\documentclass[journal]{IEEEtran}
\IEEEoverridecommandlockouts
\usepackage{graphicx, subcaption}
\usepackage{amsmath,amsthm,amsfonts,amssymb}
\usepackage{cite, soul}
\usepackage{url}
\usepackage{bm}
\usepackage{comment,enumitem}

\usepackage[usenames,dvipsnames]{color}

\theoremstyle{plain}

\title{AI-Driven Fronthaul Link Compression in Wireless Communication Systems: Review and Method Design}

\author{
Keqin~Zhang
\thanks{K. Zhang is with the school of Electrical Information and Communications, 
Huazhong University of Science and Technology, 430074, Wuhan, Hubei, China 
(E-mail: Kertsing@hust.edu.cn).}
\thanks{This manuscript is a course assignment for the 2025 AI Summer School at Huazhong University of Science and Technology. Owing to the author’s limited knowledge and experience, errors may remain; readers are advised to exercise their own judgment.}%

}

\begin{document}

\maketitle
\begin{abstract}
    Modern fronthaul links in wireless systems must transport high-dimensional signals under stringent 
    bandwidth and latency constraints, which makes compression indispensable. 
    Traditional strategies—such as compressed sensing, scalar quantization, and fixed-codec pipelines—often 
    rely on restrictive priors, degrade sharply at high compression ratios, and are hard to tune across 
    channels and deployments. Recent progress in Artificial Intelligence (AI) has brought end-to-end learned 
    transforms, vector and hierarchical quantization, and learned entropy models that better exploit the 
    structure of Channel State Information (CSI), precoding matrices, I/Q samples, and LLRs.
    
    This paper first surveys AI-driven compression techniques and then provides a focused analysis of 
    two representative high-compression routes: CSI feedback with end-to-end learning and Resource Block 
    (RB)–granularity precoding optimization combined with compression. Building on these insights, 
    we propose a fronthaul compression strategy tailored to cell-free architectures. The design targets 
    high compression with controlled performance loss, supports RB-level rate adaptation, and enables 
    low-latency inference suitable for centralized cooperative transmission in next-generation networks.
\end{abstract}


\section{Introduction}

{The architecture of mobile communication networks is evolving toward centralization and intelligence. 
Cloud Radio Access Networks (C-RAN) improve spectral efficiency and coordinated processing by 
centralizing baseband processing in a Baseband Unit (BBU), but they rely on high-speed 
fronthaul link(FL) to connect distributed Remote Radio Units (RRUs)\cite{CRAN1,CRAN2}. 
Especially in massive multiple-input multiple-output (massive MIMO)\cite{MIMO} 
systems and the forthcoming cell-free massive MIMO (CF-mMIMO) networks\cite{CFmimo}, 
each time slot must convey either large-array in-phase/quadrature (I/Q) samples or 
precoding-matrix data over the FL\cite{FL}, creating enormous traffic. Without effective 
compression, fiber FL will be overwhelmed, constraining network scalability.}

Legacy fronthaul interfaces such as the Common Public Radio Interface (CPRI) \cite{CPRI1}
transmit uncompressed time-domain samples at extremely high rates 
(for example, an 8×8 MIMO signal with 20 MHz bandwidth can require tens of gigabits per second). 
To curb fronthaul load, the industry has introduced enhanced CPRI (eCPRI), which reduces 
traffic via protocol optimization and data compression\cite{CPRI2}. However, fixed signal-processing 
and compression pipelines struggle to balance performance and compression ratio across 
diverse scenarios\cite{Quantization1}. As 5G/6G services impose ever-tighter fronthaul bandwidth and latency 
requirements, intelligent, adaptive compression has become a research focused \cite{Quantization2}. 
On the one hand, high compression ratios are needed to alleviate bandwidth pressure; 
on the other, compression distortion must be kept under control so that system 
metrics—such as bit-error rate and throughput—remain within acceptable limits.

\begin{figure}
    \centering
    \includegraphics[width=0.48\textwidth]{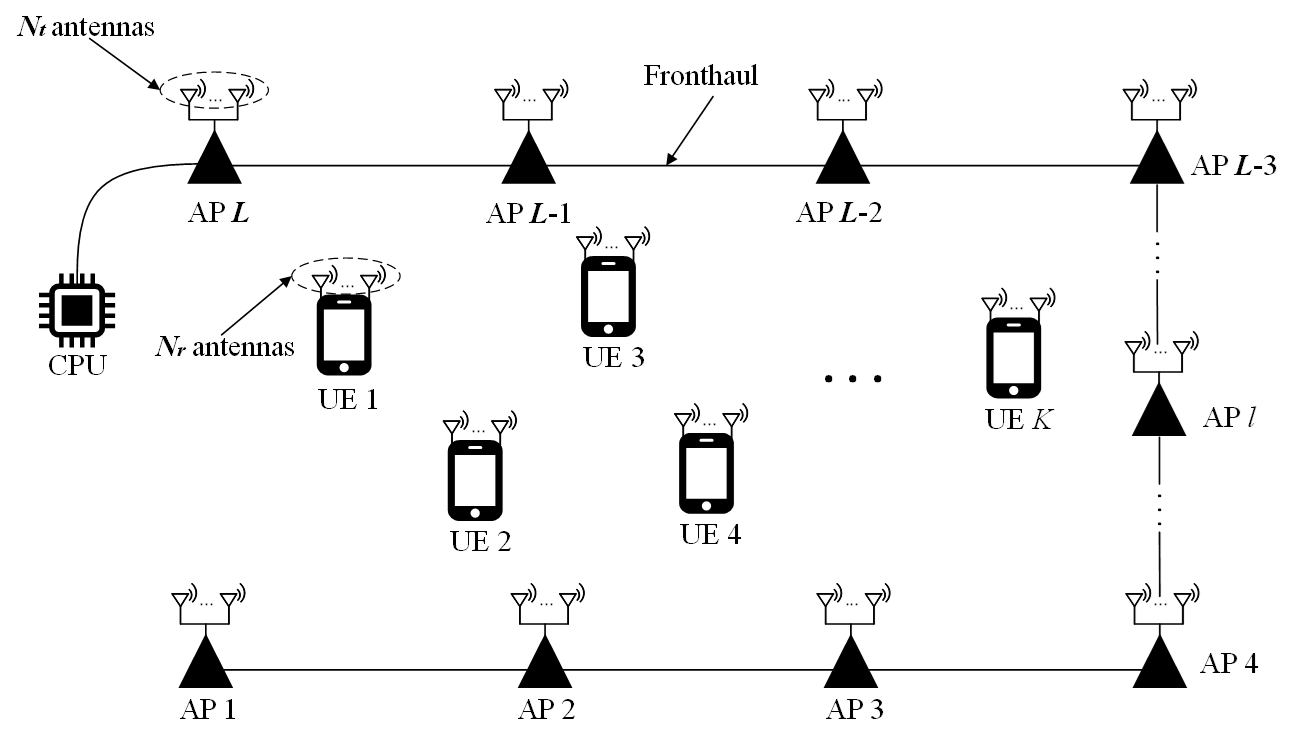}
    \caption{Advanced CF-mMIMO communication system}
\end{figure}

\begin{figure*}
    \centering
    \includegraphics[width=0.8\textwidth]{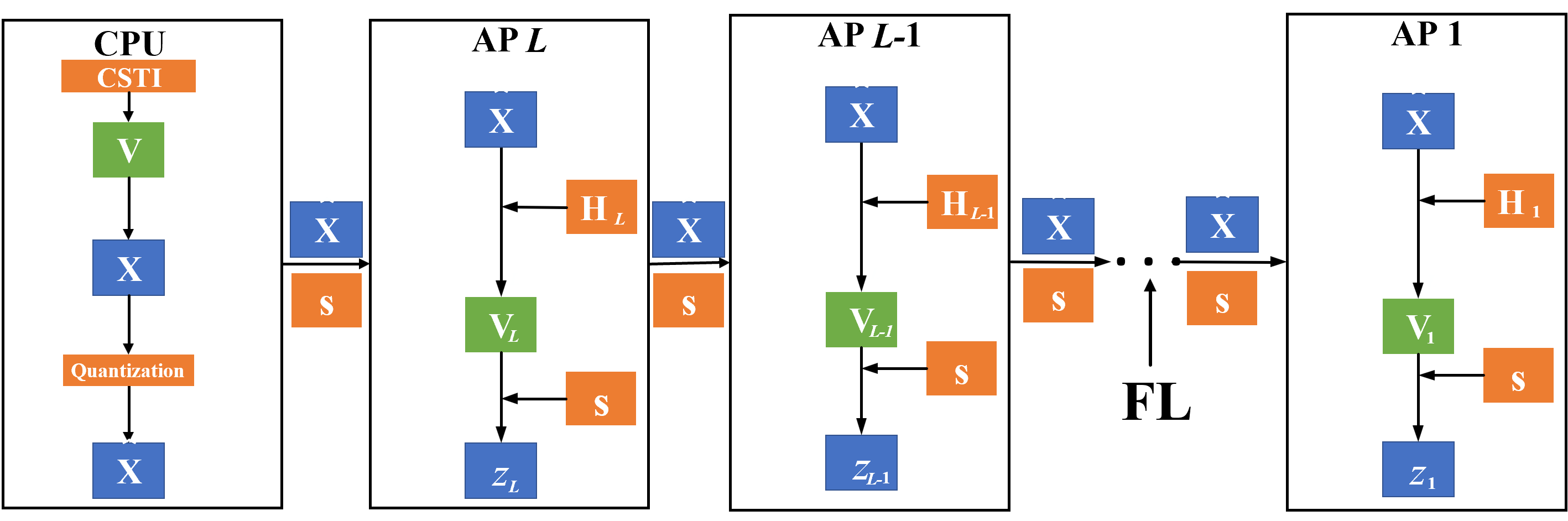}
    \caption{A compression strategy for the FL of the CF-mMIMO system\cite{CFmimo}}
\end{figure*}
The current landscape of fronthaul compression faces three major challenges:
\begin{enumerate}[leftmargin=0.45cm]
    \item Efficient representation of high-dimensional data: channel state information(CSI) matrices in massive MIMO, 
    multi-carrier I/Q symbols in FDMA, and decoder-output og-likelihood
    ratios(LLRs) are all high-dimensional 
    and highly correlated, calling for compression mechanisms that decorrelate and compactly 
    represent these signals. Traditional per-sample (scalar) quantization fails to exploit 
    inter-sample correlation, limiting compression efficiency.
    \item Performance impact of compression distortion: Compression is lossy; 
    excessive distortion reduces post-demodulation SINR, increasing BER or lowering capacity. 
    Wireless systems typically require the loss to stay within bounds—for example, 
    at most 0.5 dB SINR degradation and no more than three percent reduction in spectral 
    efficiency—imposing stringent demands on reconstruction accuracy.
    \item Real-time operation and deployment complexity: Fronthaul volumes are massive under tight 
    latency budgets, so algorithms must be highly efficient. Moreover, RRUs have limited compute, 
    hindering deployment of overly complex models. 
    Balancing compression ratio, reconstruction fidelity, and complexity remains the core challenge.
\end{enumerate}

\section{Related Works}

Conventional fronthaul compression methods include compressed sensing\cite{CS} and vector quantization\cite{quantization3}. 
Early approaches exploited signal redundancy—for example, leveraging the spectral redundancy of 
multicarrier Orthogonal Frequency-Division Multiplexing(OFDM) signals to downsample and filter, thereby lowering the sampling rate, and then 
applying nonuniform quantization to reduce the number of bits. Subsequent work introduced vector 
quantization, mapping blocks of samples to codebook indices and augmenting this with entropy coding 
to further boost compression efficiency. Although these methods improve performance to some extent, 
they still struggle to adapt to the complex, time-varying statistical characteristics of wireless signals\cite{quantization4}.

Recent advances in deep learning have opened new avenues for fronthaul compression\cite{DL1,DL2,DL3,DL4}. 
End-to-end neural models can automatically learn task-adaptive compact representations. 
In CSI feedback, the pioneering CsiNet by Wen et al. employed an autoencoder to compress and 
return downlink CSI in massive MIMO, substantially reducing overhead\cite{CSINET}. 
A subsequent wave of Deep Learning(DL) based designs incorporated convolution and attention mechanisms to 
further improve reconstruction accuracy. Nevertheless, several early DL approaches perform poorly 
in outdoor deployments at high compression ratios, and most methods optimize proxy errors 
such as normalized mean-squared error (NMSE) rather than directly constraining system-level performance, leaving their impact on 
metrics like SINR and throughput insufficiently controlled\cite{DL5,DL6}.

Within the C-RAN setting, Qiao et al. propose a learnable linear-transform compressor: 
each RRU trains a dimension-reduction matrix followed by uniform quantization to shrink 
both uplink and downlink dimensions\cite{C-RAN_DL}. To curb frequent signaling, they further employ meta-learning
with a gated recurrent unit (GRU)-based controller to reduce the number of CSI-gradient exchanges between the BBU and RRH,
 accelerating convergence and improving compression gains\cite{FL_DL}. Complementary studies account for fronthaul 
 delay and packet-loss constraints, using deep reinforcement learning to adapt the compression policy 
 on the fly. On another front, Bian et al. introduce two AI-native CPRI compressors: (i) a nonlinear 
 transform paired with learned-codebook vector quantization and (ii) an entropy-model path that maps 
 network outputs into variable-length bitstreams. They also design weight-sharing models and progressive 
 refinement to suit multi-link operation and memory-constrained deployments. Collectively, these efforts
  show that DL-based compressors can markedly enhance fronthaul efficiency and robustness, and they have
   become a focal point of current research.

Despite this progress, important gaps remain. Reconstruction accuracy still degrades at extreme compression\cite{ONEBIT}; 
many schemes do not directly optimize wireless-level metrics, leading to noticeable throughput drops; 
and some models are too heavy to deploy at the RRU. This motivates AI-driven architectures that jointly 
achieve high compression ratios, low distortion, and low complexity. Against this backdrop, 
we review two representative trajectories: (A) high-compression CSI feedback \cite{CSINET}; (B) RB-granularity 
precoding optimization plus compression (RB-WMMSE + TVQ-VAE)—and, building on their insights, 
propose an improved fronthaul scheme\cite{RB}.

\section{Review of Method A}

Method A refers to the work by Yangyang Zhang et al. (IEEE WCL 2021) \cite{CSINET}
on deep learning–based CSI compression and quantization for high 
compression ratios in FDD massive MIMO. The method targets downlink CSI feedback 
in FDD: the UE compresses the downlink channel matrix and feeds it back to the BS. 
Noting that conventional compressors suffer pronounced reconstruction degradation 
in outdoor, large-scale-fading scenarios under aggressive compression, 
Method A upgrades the CsiNet framework to CsiNet+DNN, substantially improving 
CSI reconstruction at very high compression ratios while preserving end-to-end 
learnability.

Concretely, Method A augments the decoder’s RefineNet with two fully 
connected layers to compensate for the limited ability of pure convolution 
to capture global dependencies and residual structure—particularly advantageous 
under an extremely narrow bottleneck—and replaces ReLU with Swish to improve 
trainability and stability in deeper networks by mitigating gradient attenuation 
and strengthening nonlinearity. These two modifications keep the overall 
compress–reconstruct paradigm intact yet deliver higher reconstruction quality 
under high compression and complex outdoor channels, offering a practical 
architectural and training path for subsequent DL-based CSI feedback research.

\subsection{Strong Points of Method A}

Under aggressive compression, the method attains substantial gains in 
reconstruction accuracy. Simulation results indicate that, relative to 
CsiNet plus and compressed sensing baselines, the improved CsiNet 
architecture yields higher CSI fidelity and lower NMSE in outdoor macrocell 
scenarios. Even at a compression ratio of $1/32$, the model preserves effective 
CSI reconstruction, whereas CsiNet-based baselines exhibit pronounced 
degradation. In addition, the method demonstrates robustness to 
channel-estimation noise: when estimation errors are present, 
reconstruction accuracy remains high, indicating strong noise resilience 
suitable for practical wireless environments.

From a modeling perspective, the approach is technically well-founded. 
By combining residual connections with deep fully connected layers and adopting 
the Swish activation, it enhances representational capacity and trainability, 
thereby broadening the design space for CSI compression networks and informing 
subsequent designs. From an application standpoint, the method directly addresses
 the heavy CSI feedback burden in Frequency Division Duplexing (FDD)
 systems; in 5G/6G large-antenna FDD deployments, reducing feedback overhead 
 is of clear practical value, underscoring the engineering relevance of the 
 approach.

\subsection{Weak Points of Method A}

Despite its strong performance, the method has several limitations. 
An obvious problem is the high complexity of the model.
 To attain stronger reconstruction capability, the CsiNet-based design 
 introduces deeper convolutional stacks together with a deep fully connected 
 module on the decoder side, which substantially increases parameter 
 count and computational cost; subsequent evaluations generally indicate that, 
 while accuracy improves, compute and memory usage also rise, and deployment 
 overhead is higher than lightweight Transformer-based designs—posing challenges 
 for resource-constrained UEs. 
 
 Second, end-to-end quantization optimization is insufficient. 
 Although quantization is employed, the training pipeline does not explicitly 
 address the non-differentiability of the quantizer or systematically 
 optimize quantization noise. Prior studies show that differentiable 
 approximations (e.g., $\mu$-law) or explicit quantization-error compensation 
 modules can enhance feedback performance. By contrast, the scheme resembles a 
 largely post-hoc quantization step, which may hurt reconstruction accuracy 
 when the feedback bit budget is tight.

Finally, generalization and system-level objectives require strengthening. 
While the network performs well in outdoor macrocell scenarios, 
its effectiveness across different channel models (e.g., indoor, millimeter-wave) 
and antenna scales still needs verification. Because training is 
distribution-specific, robustness depends on the coverage of the training set, 
and distribution shift can degrade reconstruction quality. In addition,
 optimization focuses primarily on proxy metrics such as NMSE, 
 without directly incorporating downstream performance after precoding—such 
 as rate loss $\Delta R$ and post-processing SINR—into the training and 
 evaluation loop. Although reductions in NMSE often correlate with throughput 
 gains, the lack of explicit constraints on system-level metrics can lead to 
 misalignment between training objectives and end performance.

 \subsection{Detailed Analysis of Method A}
 Method~A entails nontrivial computational and deployment cost. 
 The deeper decoder with stacked convolutions and fully connected 
 layers markedly increases parameter count and multiply--accumulate operations, 
 which raises memory footprint, inference latency, and energy consumption.
  Such overhead can be prohibitive for resource-constrained UEs and tight 
  real-time fronthaul budgets. Moreover, the design does not natively include 
  complexity control (e.g., structured pruning or distillation) nor
   quantization-aware training throughout, so the accuracy--latency--bit-width 
   trade-off at deployment may be suboptimal when feedback bits are limited.

 The approach is also strongly data dependent, and its generalization is not 
 guaranteed. Training is tailored to a specific channel distribution; under 
 distribution shift---across scenarios (indoor, mmWave), antenna scales, 
 mobility patterns, or hardware impairments---the reconstruction quality can 
 degrade unless the model is retrained or adapted. In addition, the optimization
  focuses on proxy errors such as NMSE rather than directly constraining 
  system-level objectives (e.g., post-precoding SINR or rate loss $\Delta R$), 
  so improvements in the training metric may not consistently translate to 
  downlink throughput gains. Overall, the method's sensitivity to the training 
  data and limited cross-scenario robustness remain important limitations 
  alongside its computational footprint.

\section{Review of Method B}

Method~B (Z. Chen \emph{et al.}, 2024; published in IEEE TNSE 2025)\cite{RB} addresses 
fronthaul compression for the cell-free architecture. 
The central idea is to optimize and compress downlink precoding at the 
\emph{Resource Block (RB)} granularity so as to reduce the fronthaul 
load from the \emph{Central Processing Unit (CPU)} to distributed 
\emph{Access Points (APs)}. In cell-free systems, the CPU computes 
joint precoders across all APs for each user, and each AP must obtain 
its corresponding precoding vectors. If precoders are designed per subcarrier 
and transmitted individually, the resulting fronthaul payload becomes 
prohibitive.

The work first develops an RB-level precoding optimization algorithm. 
The compression-aware design is cast as a stochastic, 
nonconvex program that jointly optimizes the precoder while controlling 
capacity loss induced by subsequent compression. Through analysis, 
the authors show that the RB-optimal precoding matrix lies in the column 
space of the corresponding RB frequency-domain channel, i.e., the precoding 
vectors can be expressed as linear combinations of subcarrier-wise channel 
responses within the RB. Leveraging this structure, an RB-level 
\emph{Weighted Minimum Mean Square Error (RB-WMMSE)} procedure  is 
derived to iteratively compute the Karush--Kuhn--Tucker (KKT) stationary 
solution per RB, with frequency-dependent weights to maximize the global sum 
rate. Building on the optimized precoders, the compression subproblem is 
addressed using a Transformer-based \emph{vector-quantized variational 
autoencoder (TVQ-VAE)} architecture : a Transformer encoder extracts 
global dependencies across antennas and frequencies, the bottleneck performs 
vector quantization via a learned codebook to enable bit-level representation, 
and a decoder reconstructs the precoding matrix. The variational formulation 
enables training by maximizing the evidence lower bound, and an entropy model 
with an autoregressive probability estimator plus arithmetic coding maps 
codewords to near-optimally compressed bitstreams, improving fronthaul 
efficiency relative to convolutional or fully connected autoencoder baselines.

\subsection{Strong Points of Method B}

Method~B introduces an innovation that fuses optimization theory with deep 
learning for cell-free fronthaul compression and exhibits several advantages. 
First, it markedly reduces fronthaul overhead while keeping performance 
loss controlled. By operating at the RB level, the fronthaul payload after 
compression is reduced by orders of magnitude relative to per-subcarrier 
transmission; aided by the optimization step, the network sum rate is largely
 maintained or even improved. Simulations show that, compared with conventional 
 RB-wise precoding without distortion modeling, RB-WMMSE achieves up to a 
 $101\%$ increase in sum rate; even when compression error is included, 
 combining RB-WMMSE with TVQ-VAE yields up to a $106\%$ gain over standard 
 autoencoder-based compression without the optimization stage. These results 
 indicate net performance gains at high compression ratios: with proper design, 
 compression does not induce severe degradation and may yield improvements due 
 to optimized precoding. Second, the approach is technically novel, being among 
 the first to integrate a Transformer with a VQ-VAE for matrix compression in 
 the fronthaul setting, thereby surpassing conventional autoencoders in 
 representational power and compression efficiency.

Third, the scheme is tailored to the characteristics of cell-free systems. 
RB-level processing aligns with frequency-selective channels where adjacent 
subcarriers exhibit correlation, and the Transformer encoder is well-suited to
 model complex dependencies across antennas and RBs, which explains its 
 superiority over generic methods. Fourth, the architecture is naturally 
 compatible with Cloud-RAN and cell-free deployments: precoding and compression 
 are executed centrally, and each AP only decodes its local precoding vectors, 
 imposing minimal additional requirements on AP hardware and facilitating 
 distributed rollout. Finally, by using sum rate as a system-level objective 
 directly within the optimization, the scheme ensures practical utility; 
 in contrast to methods that optimize solely for reconstruction error, it 
 targets end performance, which is an important direction for future AI-based 
 compression.

 \subsection{Weak Points of Method B}
 Although Method~B demonstrates strong performance, 
 several practical limitations merit attention. 
 First, the algorithmic complexity is nontrivial. 
 RB-WMMSE requires iterative optimization whose cost scales 
 with the numbers of users and APs; TVQ-VAE embeds a Transformer 
 and a vector-quantization codebook, and its inference entails 
 self-attention computation and probabilistic (entropy) coding, 
 leading to notable compute and memory overheads. 
 In large-scale cell-free networks, executing full optimization and 
 deep compression every frame may challenge real-time constraints. 
 Mitigations include lengthening the precoder update period 
 (e.g., updating every few frames) and exploiting parallel hardware 
 acceleration for Transformer inference.

Second, issues of model training and generalization arise. 
TVQ-VAE typically requires extensive offline training on representative 
data and joint optimization of the VAE and the codebook; the trained model 
depends on channel statistics and network scale, so changes in AP/user 
counts or antenna configurations may necessitate retraining or codebook 
retuning. Moreover, as sequence length grows (e.g., when matrices are flattened),
 Transformer training can be constrained by memory and data requirements. 
 Third, latency and synchronization must be considered: Method~B assumes 
 global CSI at the CPU for precoding and subsequent compression, yet CSI 
 estimation, feedback, and computation incur delay, which can reduce precoding 
 efficacy under high mobility. Fourth, implementation complexity increases 
 relative to traditional pipelines: APs must host a VAE decoder, and the central 
 unit must run deep inference together with WMMSE, potentially requiring 
 ASIC/FPGA accelerators as current baseband stacks lack mature support for 
 such models. Finally, the scheme is tailored to downlink precoding; extending 
 it to uplink (e.g., AP-side compression of received signals to the CPU) would 
 require redesign to reflect uplink data characteristics.

 \subsection{Detailed Analysis of Method B}

The RB-WMMSE stage requires iterative updates whose cost grows 
with the number of users $K$, the aggregate number of AP antennas $M$, 
and the RB count $|G|$; per-iteration matrix factorizations/inversions 
and inter-RB weighting introduce a superlinear scaling that stresses 
real-time budgets in large cell-free deployments. The TVQ-VAE stage 
further incurs Transformer self-attention complexity $\mathcal{O}(L^{2})$ 
with sequence length $L$ (e.g., the flattened precoding matrix tokens), 
nontrivial codebook memory for vector quantization, and serial latency 
from autoregressive entropy coding. End-to-end, meeting tight fronthaul 
deadlines every frame can be challenging; amortized updates (e.g., updating 
every few frames) mitigate load but reduce tracking fidelity under mobility. 
Centralized execution also makes performance sensitive to CSI staleness and 
fronthaul jitter.

The learned compressor is trained offline and is tied to the channel statistics,
 array geometry, and network scale seen during training; 
 domain shift across scenarios (indoor vs.\ outdoor, sub-6\,GHz vs.\ mmWave), 
 mobility patterns, hardware impairments, or changes in $K$, $M$, and $|G|$ 
 can degrade reconstruction unless the model and its codebook are retuned. 
 The mismatch between the compression-error model assumed in RB-WMMSE and the 
 actual error induced by the learned quantizer may bias the optimization, and 
 variable-length bitstreams are fragile to packet loss without joint 
 rate-control and error-protection design. Finally, embedding deep models 
 (Transformer, VQ-VAE) into the baseband pipeline raises implementation 
 complexity (memory footprint, accelerator requirements) and may limit 
 portability across heterogeneous radio platforms.

 \begin{figure*}
    \centering
    \includegraphics[width=0.99\textwidth]{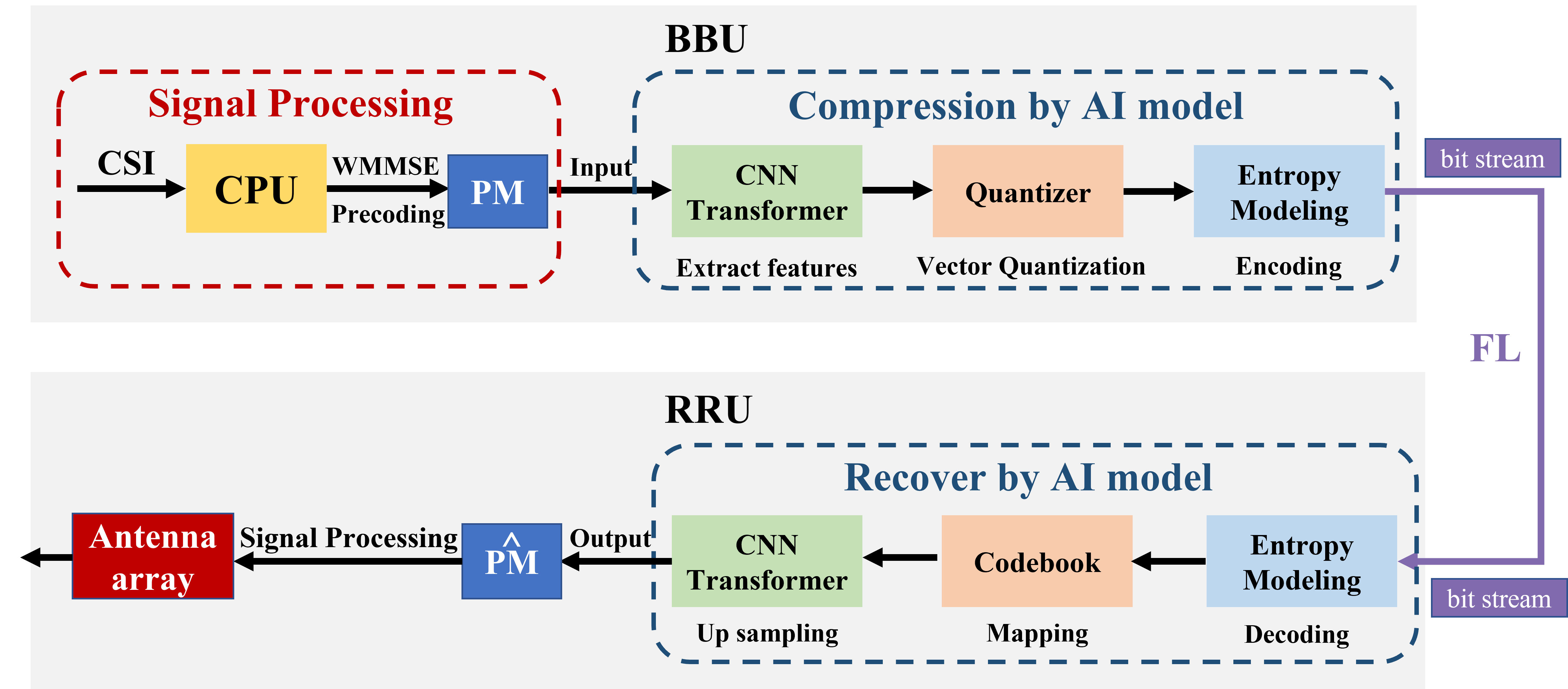}
    \caption{The FL compression strategy based on AI model proposed in this article}
    \end{figure*}

\section{Proposed Method of This Paper}
\subsection{Overall Design}
Building on the above, 
an end-to-end neural compression framework for efficient FL 
compression of downlink precoding matrices(PMs) in C\mbox{-}RAN method is proposeed in this paper.
The framework comprises an analysis transform, quantization and coding, an optional entropy model,
 and a synthesis transform. 
 The input is a set of frequency-domain PMs over multiple RBs 
 (complex entries split into real and imaginary channels); 
 the output is a compact, reconstructible codeword sequence. 
 Trained end-to-end, this framework supports RB-granular, 
 flexible compression with low inference latency, 
 matching the centralized cooperative transmission requirements of 6G cell-free deployments.


\subsection{End-to-End Neural Framework for Precoding Compression}

\paragraph{Learnable Analysis Transform}
We use an analysis transform to extract features and reduce the dimensionality of the precoding matrices (PMs). RB-wise PMs are stacked as $\tilde{\mathbf{V}}\in\mathbb{R}^{D\times 2N_t}$ by splitting real and imaginary parts, and the encoder maps them to a latent
\begin{equation}
\mathbf{z}\;=\;f_{\theta}(\tilde{\mathbf{V}})\in\mathbb{R}^{T\times d}.
\end{equation}
Residual CNN blocks describe local structure, while a lightweight self-attention layer captures long-range dependence. The second dimension is decimated after several blocks, e.g.,
\begin{equation}
D\times 2N_t \;\rightarrow\; D\times N_t \;\rightarrow\; D\times \tfrac{N_t}{2} \;\rightarrow\; \cdots \;\rightarrow\; T\times d,
\end{equation}
so the latent is compact yet informative across RBs. The module is easy to tune: depth, width, number of attention heads, and the downsampling ratio can be adjusted to meet a chosen complexity or compression budget.

 \paragraph{Vector Quantization and Hierarchical Quantization}
 The latent is discretized by learnable vector quantization (VQ) with codebook $\mathcal{E}=\{\mathbf{e}_k\}_{k=1}^{K}\subset\mathbb{R}^{d}$. Each token $\mathbf{z}_i$ is mapped to its nearest codeword,
 \begin{equation}
 c_i \;=\; \arg\min_{k}\bigl\|\mathbf{z}_i-\mathbf{e}_k\bigr\|_2^{2},
 \qquad
 \hat{\mathbf{z}} \;=\; \bigl[\mathbf{e}_{c_1},\ldots,\mathbf{e}_{c_T}\bigr].
 \end{equation}
 To support adaptive bitrate, we use $L$ refinement stages with residual updates:
 \begin{align}
 \mathbf{r}^{(0)} &= \mathbf{z},\qquad
 \mathbf{c}^{(l)} = Q^{(l)}\!\bigl(\mathbf{r}^{(l)}\bigr),\qquad
 \hat{\mathbf{r}}^{(l)} = \mathcal{E}^{(l)}\!\bigl(\mathbf{c}^{(l)}\bigr), \nonumber\\
 \mathbf{r}^{(l+1)} &= \mathbf{r}^{(l)} - \hat{\mathbf{r}}^{(l)},\qquad
 \hat{\mathbf{z}} \;=\; \sum_{l=0}^{L-1}\hat{\mathbf{r}}^{(l)}.
 \end{align}
 Training follows the VQ-VAE straight-through scheme with codebook and commitment terms.

 \paragraph{Entropy Modeling}
 An optional entropy model assigns probabilities to the indices $c_i$ (possibly with context), enabling near-entropy coding. The expected code length is
 \begin{equation}
 \mathbb{E}[L] \;\approx\; -\sum_{i}\log_{2} p_{\phi}\!\bigl(c_i \mid \text{ctx}_i\bigr)
 \;\triangleq\; H(C), 
 \end{equation}
 \begin{equation}
    R_{\text{total}} \;=\; \frac{H(C)}{T} \;+\; \sum_{l=0}^{L-1} R_l,
 \end{equation}
 A fronthaul budget is enforced by activating a subset of refinement stages or by tuning the rate weight so that $R_{\text{total}}\!\le\!B_{\text{fh}}$.

 \paragraph{Decoder}
 The decoder mirrors the encoder: it performs codebook lookups to recover the latent $\hat{\mathbf{z}}$ and then applies a synthesis network with upsampling and inverse attention/convolution to reconstruct the PMs,
 \begin{equation}
 \hat{\mathbf{V}} \;=\; g_{\psi}(\hat{\mathbf{z}}).
 \end{equation}
 Reconstruction fidelity is measured by the mean-squared error for an $M\times N$ matrix,
 \begin{equation}
 \mathrm{MSE} \;=\; \frac{\|\hat{\mathbf{V}}-\mathbf{V}\|_{F}^{2}}{MN},
 \end{equation}
 and training minimizes a closed-loop rate–distortion objective
 \begin{equation}
 \mathcal{L} \;=\; D_{\text{MSE}} + \lambda\, R_{\text{total}} \;+\; \gamma\,\mathcal{L}_{\text{VQ}},
 \end{equation}
 with optional system alignment (e.g., adding a term that penalizes the loss of downlink sum-rate). This ties analysis, quantization, entropy coding, and synthesis to the final communication performance while respecting the fronthaul budget.

 \begin{figure}
    \centering
    \includegraphics[width=0.45\textwidth]{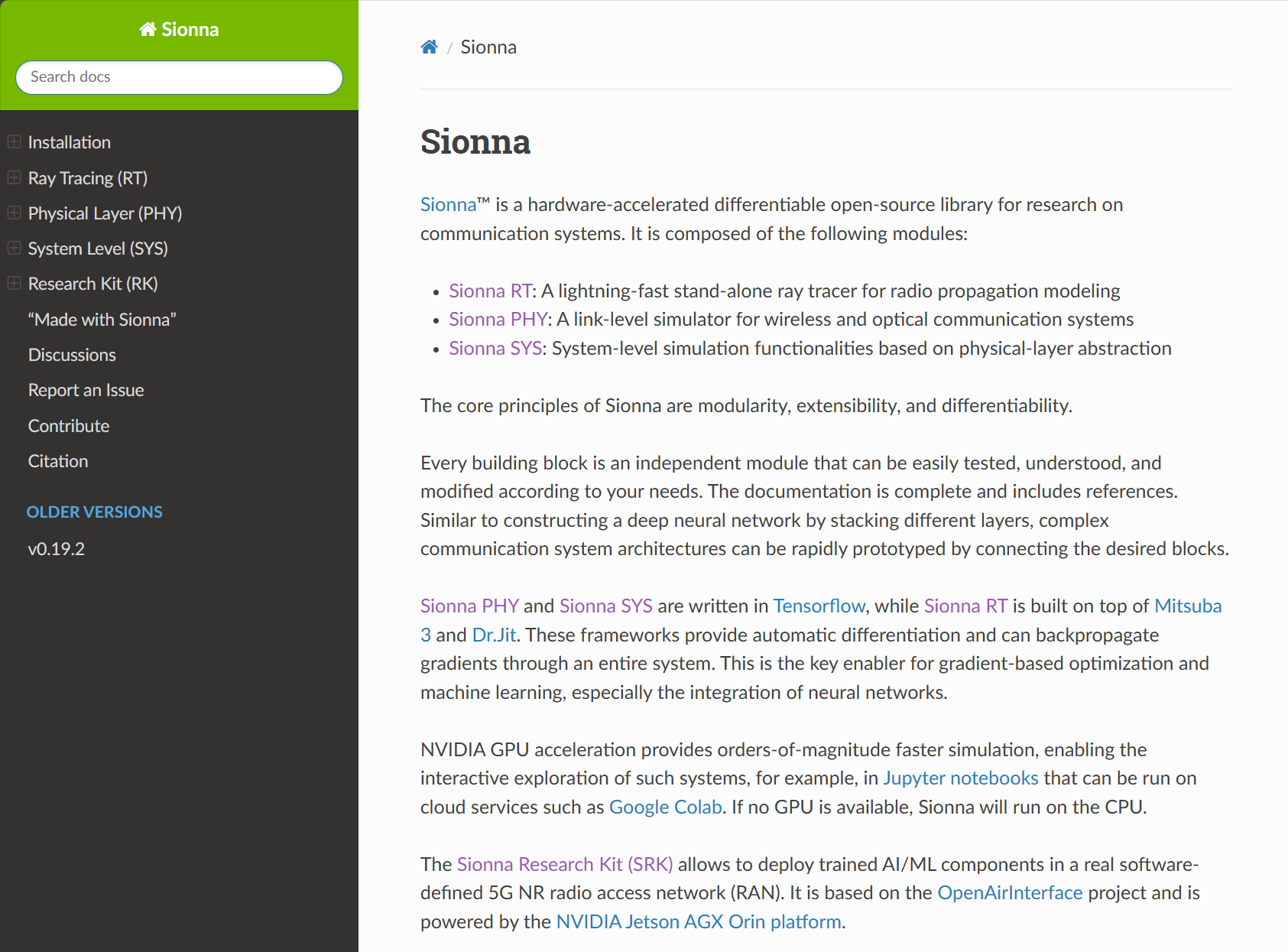}
    \caption{Documents of Sionna}
    \end{figure}

\subsection{Dataset Collection}

The AI compression module is trained on a precoding dataset generated with \textit{Sionna}, 
NVIDIA's open-source 6G physical-layer simulator\cite{sionna}. 
Sionna supports end-to-end differentiable modeling and GPU 
acceleration across multiple devices, enabling large-scale parallel generation of independent 
channels. It implements a broad set of propagation models from \textit{3GPP TR~38.901}, including 
\textit{Tapped Delay Line (TDL)} and \textit{Clustered Delay Line (CDL)} profiles for urban 
macro/micro and rural macro scenarios, and can output frequency-domain responses at the 
RB-granularity. In this work, we simulate directly in the frequency 
domain by defining an OFDM resource grid and extracting per-RB channel matrices.

As a representative cell-free configuration, we consider a carrier at $3.5\,\mathrm{GHz}$, 
delay spread of $800\,\mathrm{ns}$, and a $20$-path CDL channel. Multiple 
\textit{Distributed Units (DUs)} and user terminals are deployed; 
each DU employs a dual-polarized $8\times 16$ antenna panel communicating with users under 
frequency-selective fading. Using Sionna, we generate large numbers of independent channel 
impulse responses and, for each realization, compute RB-wise downlink precoders via the 
WMMSE algorithm, thereby constructing a “large-scale MIMO precoding-matrix” 
dataset for training. GPU acceleration is used throughout the pipeline 
(IFFT/FFT, channel generation, and precoding), allowing rapid dataset construction
 while preserving simulation fidelity.

\section{Discussions and Future Works}

This paper first reviews fronthaul compression for C\mbox{-}RAN and CF\mbox{-}mMIMO. The line of research has moved from classical tools—compressed sensing and vector quantization—to learned schemes such as the CsiNet family, attention-based encoders, and entropy–model–driven coders. We organize the literature by data type (CSI, PMs, I/Q samples, and LLRs) and by evaluation axes: compression ratio, reconstruction error, system-level performance, algorithmic complexity, and ease of deployment. Two representative paths are contrasted: (i) a CSI-feedback track that targets very high compression; and (ii) an RB-level precoding track that couples optimization with compression. A common trend is toward communication-aware training objectives, RB-granular processing that exploits frequency selectivity, and layered quantization with entropy coding that approaches the entropy bound.

Building on these observations, we develop an end-to-end neural fronthaul compressor. A learnable analysis transform reduces PM dimensionality and extracts structure using residual CNN blocks for local patterns and a lightweight self-attention module for long-range dependence. Vector quantization produces discrete indices; a base–refinement hierarchy yields an adaptive bitstream. An optional entropy model further compresses the index within its entropy value range. A synthetic decoder reconstructs the PM (pre-encoding matrix). Training uses the MSE term, and may include system-level objectives such as the total downlink rate and the EVM. This design operates at the RB level, supports flexible bit rate control, and is optimized for low inference latency for centralized cooperative transmission suitable for the 6G environment. 

This method still has limitations. The computation and latency must be strictly managed: attention and entropy encoding will increase the CPU load, while the codebook storage combined with the nearest neighbor search will introduce additional overhead. Performance may depend on channel statistics, antenna/user scale, and the staleness of channel state information, which affects generalization ability. Variable-length bit streams also require joint rate control and error prevention mechanisms.

\bibliographystyle{IEEEbib}
\bibliography{abs_ref}

\end{document}